# Low Fidelity VTOL UAV Design Optimization Using an Open Source Framework

Nikhil Sethi
*Mechanical Engineering Undergraduate*
*(Delhi Technological University)*

Saurav Ahlawat
*Applied Physics Undergraduate*
*(Delhi Technological University)*

Prof. N. S. Raghava
*HoD ECE Department*
*(Delhi Technological University)*

## Nomenclature

| | | | | |
|---|---|---|---|---|
| $W$ | Weight | | $P$ | Power |
| $T$ | Thrust | | $ROC$ | Rate of Climb |
| $GTOW$ | Gross Take-off Weight | | $FOM$ | Figure of Merit |
| $C_D$ | Coefficient of drag | | $S$ | Wing reference area |
| $C_L$ | Coefficient of lift | | $\eta$ | Efficiency |
| $C_M$ | Pitching moment | | $\mu$ | - |
| $\rho$ | Density of air | | $\theta$ | Motor tilt angle |
| $\sigma$ | Areal density | | $K$ | - |
| $W/S$ | Wing loading | | $_{vt}$ | Vertical tail components |
| $W/P$ | Power loading | | $_{ht}$ | Horizontal tail components |
| $L$ | Lift (N) | | $_{vtol}$ | VTOL assisting components |
| $D$ | Drag (N) | | $C$ | Rated charge capacity |

*Abstract*—An all-electric unmanned aerial system with both VTOL and Fixed wing capabilities is designed and optimized for long range surveillance and relief operations. The UAV is equipped with onboard computers and sensors and is capable of carrying 1kg of relief payload upto 100 Km. The entire low fidelity design process- from concept to render is carried out using completely open source tools, libraries and in-house code. The challenges faced and primary differences are discussed parallelly. A comparison with commercial codes and programs is also done in some areas to give an overview of key capabilities and caveats.

## I. Introduction

The development of Unmanned Aerial Vehicles is gaining popularity with the advent of more research and robust tools for the same. An even more recent development is the VTOL hybrid concept which combines the advantages of both fixed wing and vertical thrust configurations. They can adapt to the mission scenario while keeping good performance characteristics throughout. One of the primary challenges in developing such vehicles is the added weight and parasite drag of the vertical thrust components as compared to conventional designs. The low operating Reynolds number of MAVs makes this an even more difficult problem to tackle.

A key challenge that undergraduate students and inexperienced designers face is the unavailability of consolidated resources to learn and develop good designs. This is even more pronounced in the case of aircraft design where the intertwined relationships of multiple disciplines are profound. The use of commercial tools act as a proponent in such a case as they do not provide the user with the freedom to learn and customize accordingly. Opensource tools and the development of custom libraries/components helps in learning from scratch and the leverage to create and use such tools has never been greater than before.

The authors recognize the above two challenges and aim to alleviate them by developing a sound methodology spread across a wide variety of completely free tools and in-house code. Our research is aimed primarily to help students and upcoming researchers developing feasible designs for student competitions or learning to design in general. Formally, the rationale of our research is centered around the following outcomes:

Tangible: The development of a robust and stable design for a VTOL hybrid UAV within the defined design constraints.

Intangible: Inculcation of completely open source tools and libraries for the entire design and testing routine so as to move towards "free design" methodologies.

The paper is organized as follows- beginning with a background and literature review of the current state of research carried out by others. This is followed by our planning and execution explained in the Methods section and the entire design process. A brief section on the fabrication and testing is also given. To conclude, some final comments and discussion on the design and methods used is carried out.

*Note: This research extensively uses the term "hybrid" which is not to be confused with hybrid electric propulsion which relies on the use of gasoline powered generators for the production of onboard power. However, "hybrid" here would only refer to the operational capabilities of the aircraft*

as both a multirotor and a fixed wing craft unless stated otherwise.

## II. BACKGROUND

A VTOL fixed wing UAV has the ability to take off and land vertically as well as hover in place along with achieving a sustained cruised flight. These kind of hybrid UAVs combine VTOL capability with the standard forward propulsion of a fixed wing UAV. In many hybrid VTOL UAVs, rotary lift propellers are typically incorporated into the aircraft's wings, which then transition for forward flight. VTOL fixed wings have several advantages over typical fixed-wing unmanned aircraft. They require much less space to launch and recover, as they do not need to use a specified area for Take-Off. They are suited to applications where aerial inspection and monitoring is needed, making the aircraft maintain a fixed position for a period of time

A study of [1], covers the use of SUAVE, a software used to analyze, optimize, and design a range of small UAVs which has been shown to be a flexible aircraft design tool with the unique ability to handle alternative energy systems and unconventional designs. In [2], a similar configuration to this paper is explored using standard low fidelity analyses. The nature of the software and tools used is not stated however. In [3], potential design strategies are presented and compared, identifying various design variables and constraints which take precedence during a particular optimization while taking a novel weight build-up approach to estimating operating empty weight. This work simplifies the larger study of optimizing sensitivity of constraints while removing implicitly resolved parameters alongside reducing computational cost.

In this work, a wide variety of tools and resources were used in the design and development and a brief description of each and their capabilities is highlighted below:

*1) FreeCAD:* It is a powerful open source solid modelling software written in C++ with a highly interoperable Python API. It has multiple workbenches to deal with different kinds of geometry. Being a relatively old software, the documentation is sufficient and has a well-maintained forum.[1]

*2) OpenFoam+ParaView:* Another toolbox written C++ for the development of sutomized numerical solvers containing all pre as well as post-processing utilities for the solution of continuum mechanics problems in computational fluid dynamics.[1] The library is also distributed with Paraview which is a verstatile post processing software to visualise the computed fields.

*3) LuxCoreRender:* This software provide artists with a powerful tool to create extremely realistic and accurate images. It is very flexible to the user and does not compromise on quality and physical correctness where the calculations are done according to mathematical models based on physical phenomena aiming to make 'unbiased' design choices.

*4) OpenVSP:* It is a parametric aircraft geometry tool where the user is allowed to create a 3D model of an aircraft defined by common engineering parameters. This model can be processed into formats suitable for engineering analysis. This software was released as an open source project under the NASA's Open Source Agreement (NOSA) version 1.3. giving the advanced tools of aircraft design for use to the general public.[5]

*5) XFLR5:* It is an analysis tool for airfoils, wings and planes operating at low Reynolds Numbers which includes XFoil's Direct and Inverse analysis capabilities with wing design and analysis capabilities based on the Lifiting Line Theory, on the Vortex Lattice Method, and on a 3D Panel Method.

*6) Ardupilot SITL/Mission Planner:* The SITL (software in the loop) simulator allows us to run Plane and Copter simulations without any actual flights. It is a build of the autopilot code using the original C++ compiler, giving us a native executable that allows one to test the behaviour of the code without implementation on the hardware itself. The SITL uses the sensor data coming from various flight dynamics models in a flight simulator. This allows ArduPilot SITL to be tested on a very wide variety of vehicle types, essential for testing a hybrid system such as ours.

## III. DESIGN METHODOLOGY

Incorporating completely open source tools for the conceptual preliminary and detailed design is a challenge because of three main factors:

1) The accuracy of results
2) Lack of documentation
3) Lack of experienced users/populated forums

Considering the above challenges and the fidelity required for the design, the tools along with the design process are shown in fig 1.

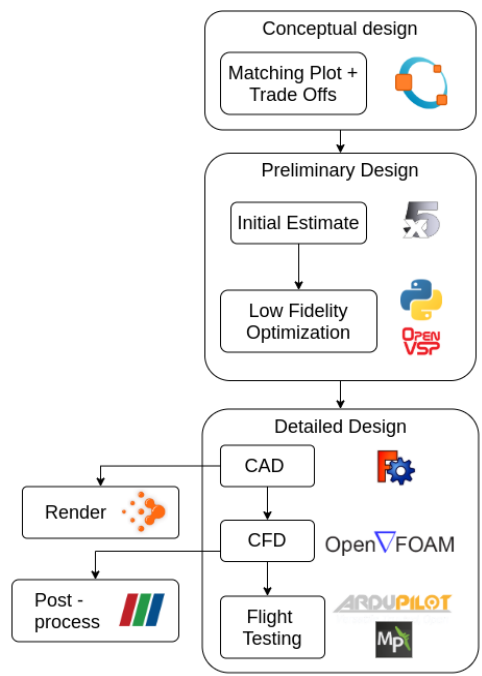

*Fig 1. High Level Design Process.*

The high degree of inter-related disciplines in aircraft design make it difficult to dodge locally optimal designs developed by intuition. Multidisciplinary design optimization has therefore become a necessary part of the toolbox of aerospace designers. The approach used in the following sections is not a full scale MDO solution by any means but a novel low fidelity attempt at design exploration. A human is kept in the loop at all stages to guide the design and perform validation checks. Further methods pertaining to the same are explained in the following sections

## IV. CONCEPTUAL DESIGN

For the purpose of this research, we model our aircraft for maximizing its range of flight to a 100kms, cruising at 25m/s while maintaining its VTOL functionality. The aircraft will keep a standard mission profile with vertical take-off, cruised flight and vertical landing.

### A. Configuration selection:

A hybrid of a blended flying wing configuration supplemented with a tricopter configuration with two forward tilting motors is arrived at. The forward motors tilt forward for transition to cruise flight after a vertical take-off and tilt back to multirotor while landing. The rear motor is operational for the multirotor segment of flight.

$$(C_{Do})_{Total} = (C_{Do})_W + (C_{Do})_{ht}\frac{S_{ht}}{S} + (C_{Do})_{vt}\frac{S_{vt}}{S} + (C_{Do})_{vtol}\frac{S_{vtol}}{S}$$

A blended flying wing will prove to be a better candidate for long endurance missions with significantly reduced $(C_{Do})_{ht}$ component and better performance of winglets to reduce vortex induced drag and an overall lower $(C_{Do})_{vt}$ component.

Most commercially available products use a quad-plane configuration, which is basically a quadcopter attached to the frame of an independent fixed wing aircraft. This solution, though the least complicated, added redundant structures and additional motors which contribute more to $(C_{Do})_{vtol}$ component for additional drag and are basically dead weight in cruised flight. This would not be an optimal choice for a high endurance mission.

A tri plane with two forward tilt motors however is a highly favorable choice in terms of weight and lower $(C_{Do})_{vtol}$. Even though planes with a greater number of rotors provide better stability and airworthiness, their cost and weight diminish the advantages. Owing to a tri copter's superior stability and airworthiness along with its cost effectiveness, it is chosen.

### B. Propulsion:

The propulsion selection is the most essential tradeoff to obtain the most out of the aircraft's endurance and reach the desired range.

During cruise flight, all the load will be on the forward motors whereas during hover, all the three motors would share the load. This introduces the issue of transition where the forward motors have to be chosen in such a way so as to provide sufficient thrust during hover as well as have optimal RPM and pitch speed for a sustained cruised flight.

The thrust margin of the all the motors was arrived at an $x + x + 2x$ ratio where the max thrust of the forward motors is capped at $x$. This distribution was a necessary step to ensure that the forward motors operate at their peak optimum during cruise and the rear motor could handle the remaining thrust required for hover. The rear motor, only operational during hover segments, is optimized for hover and the ratio of thrust distributed on it is calculated based on the available thrust margin from the forward motors after they have been finalized during selection for cruised flight.

### C. Weight Estimation:

The initial estimation is done by analyzing historical data, collecting the necessary features and arriving at a hypothesis akin to the methods in [7]. Since the payload is defined within the constraints, the payload fraction was used as the most important feature at this stage and a linear regression curve is fitted to the data. This trendline along with the equations is solved analytically to give a prediction.

$$W_0 = a\left(\frac{W_p}{W_0}\right) + b$$

$$W_0 = W_e + W_p$$

where a = - 0.00468; b = 0.878

The regression analysis yields the following total mass of the aircraft required for 1 kg of payload. A 200g safety margin brings this up to an estimated 7kg aircraft.

Payload = 1 kg

Total = 6.8 + 0.2 = 7kgs

## V. PRELIMINARY DESIGN

The preliminary design stage makes use of extensive mathematical optimization. The objective of the optimization process is to design a geometry and propulsion system to maximize the range which implies minimizing power and maximizing cruise efficiency. In our case, we have divided the optimization and design process into three stages.

### A. Initial Sizing:

The first stage gives a good estimate of the two driving factors for any design- wing loading and power loading. Appropriate relations for the transition, hover and climb phases for the tricopter sizing are also included to account for disc loading using an approach very similar to the one in [8]. Since a rough estimate is desired at this stage, a design space is created using Linear programming based on the following formulae:

$$\left(\frac{W}{S}\right)_{stall} = 0.5\rho(V_{stall})^2 C_{Lmax}$$

$$\left(\frac{W}{S}\right)_{max\,range} = q\sqrt{\pi A e C_{D_o}}$$

$$\left(\frac{W}{S}\right)_{max\,loiter} = q\sqrt{3\pi A e C_{D_o}}$$

$$\left(\frac{W}{P}\right)_{FW\ ROC} = \frac{1}{\frac{ROC}{\eta_p} + \sqrt{\frac{2}{\rho\sqrt{\frac{3C_{D_o}}{K}}}\left(\frac{W}{S}\right)}\left(\frac{1.155}{\left(\frac{L}{D}\right)_{max}\eta_p}\right)}$$

$$\left(\frac{W}{P_{SL}}\right)_{V_{max}} = \frac{\eta_p}{\frac{1}{2}\rho_o V_{max}^3 C_{D_o}\frac{1}{\left(\frac{W}{S}\right)} + \frac{2K}{\rho\sigma V_{max}}\left(\frac{W}{S}\right)}$$

$$\left(\frac{W}{P}\right)_{hover} = FOM \times \left(\frac{\sqrt{2\times\rho}}{DL}\right)$$

$$\left(\frac{W}{P}\right)_{climb} = \frac{1}{V_y - \frac{k_1 V_y}{2} + \frac{k_1}{2}\sqrt{V_y^2 + \frac{2(DL)}{\rho_o}} + \frac{\rho_o V_{tip}^3}{(DL)}\left(\frac{\rho C_d}{8}\right)}$$

Hence,

$$\left(\frac{W}{P}\right)_{trans} = \frac{1}{d_1 + d_2 + d_3}$$

$$d_1 = \frac{k_1}{\sin(\theta_{tilt})}\sqrt{\frac{-V_\infty^2}{2} + \sqrt{\left(\frac{-V_\infty^2}{2}\right)^2 + \left(\frac{DL}{2\rho\sin(\theta_{tilt})}\right)^2}}$$

$$d_2 = \frac{\rho V_{Tip}^3}{DL}\left(\frac{\sigma C_d}{8}(1 + 4.6\mu^2)\right)$$

$$d_3 = \left(\frac{1}{2}\rho_o V_\infty^3 C_{D_o}\frac{1}{\left(\frac{W}{S}\right)} + \frac{2K}{\rho_o V_\infty}\left(\frac{W}{S}\right)\right)$$

The LPP outputs a comfortable design space to choose points from depending on desired performance. To make the analysis more constrained, minimum and maximum values of disc loading are also used depending on the motors and their rated propeller sizes. As shown, the VTOL and fixed wing design points are far apart, both constrained by their climb limits. For each configuration, the maximum power points are chosen to act as a safety factor.

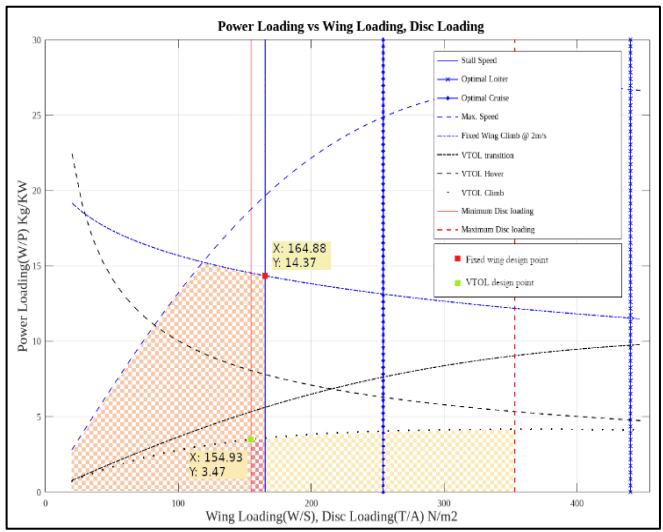

*Fig 2. Preliminary Design Space*

$$S_{wing} = \frac{W_o}{(W/S)_{opt}}$$

$$P_{max} = \frac{W_o}{\max\left\{(W/P)_{FW},\ (W/P)_{vtol}\right\}}$$

### B. Geometry optimization

#### 1) Initial 'Human' estimate

An initial estimation of aircraft performance is first analyzed on XFLR5 focusing solely to obtain initial but educated estimates of static stability parameters.

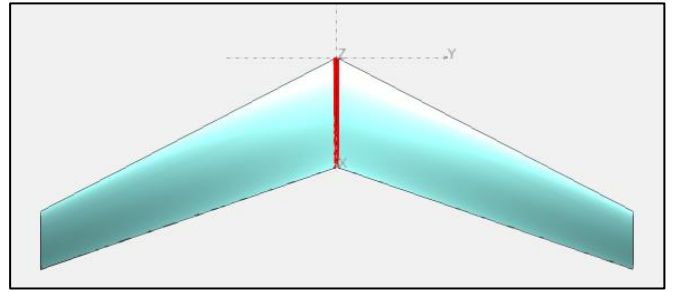

*Fig 3. Initial Wing Planform*

Wing planform area $S$ is initially estimated by using crude $C_l$ values obtained from observing trends in standard use reflexed airfoils on angle of attack at peak $C_l/C_d$ values operating at cruising altitude and speed Reynold's numbers.[10]

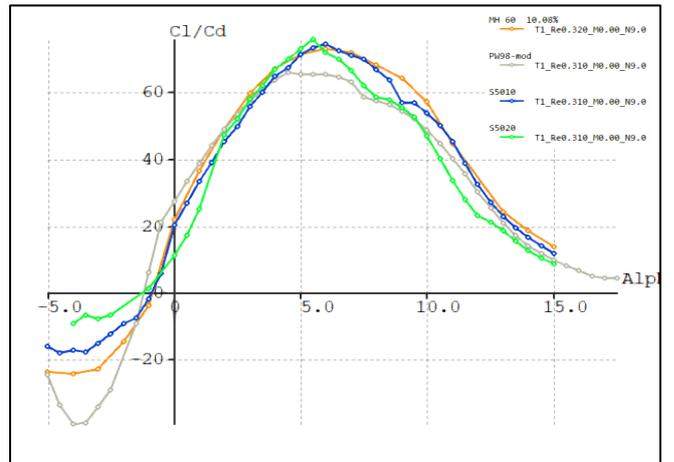

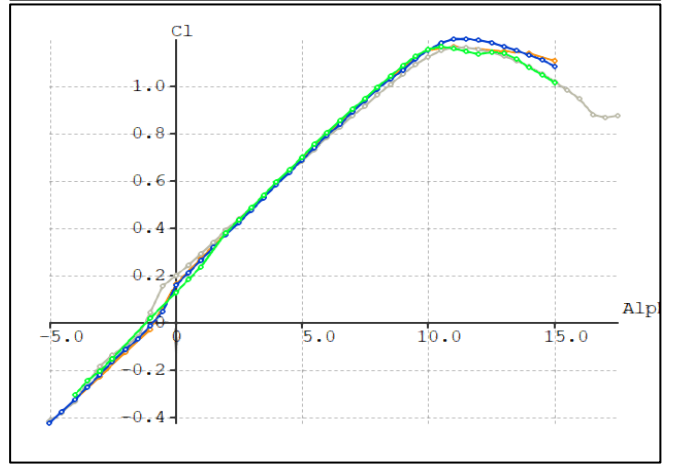

*Fig 4. Initial Airfoil Study*

Fixed lift analysis is then conducted on the operating GTOW of the flying wing now to introduce static stability into the craft by tweaking the tip twist in the wing with an optimal sweep backwards, aiming to nullify pitching moment $C_m$ of the flying wing.

Iterations were performed on the design variables to optimize the cruise $C_l/C_d$ of the whole wing while balancing out the placement of the CoG to ensure close to null pitching moment. The resultant twist in the wing subsequently reduced the $C_L$ of the whole wing, making the wing area more than initial calculations.

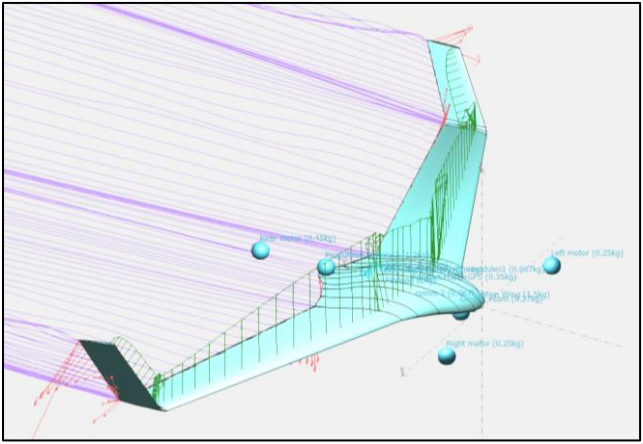

Fig 5. XFLR5 Preliminary Design

A rough and conceptual estimate of the intended design is also analysed but not optimized to ensure consistency in performance estimates.

*2) Problem Setup*

$$Min_x \quad w^T \cdot J$$
$$w^T = [w_1, w_2 \dots w_{\#J}]$$
$$J = \left[C_D, \frac{1}{n}\sum_{i=1}^{n}\left(\frac{c_l}{c_{ref}}(x)^i_{actual} - \frac{c_l}{c_{ref}}(x)^i_{target}\right)^2, \zeta_{sp}, \zeta_{dr}, \zeta_{ph}\right]$$
$$0 \leq x \leq b/2$$
$$x = [AR, \theta_{rel}, \lambda, \Lambda_{LE}]^T$$

Subject to
$$C_L = 0.9 \times C_l^c$$
$$S.M. = 4\%$$
$$C_M \approx 0$$
$$5 \leq AR \leq 10$$
$$-1 \leq \theta_{rel} \leq -5$$
$$0.4 \leq \lambda \leq 0.9$$
$$20 \leq \Lambda_{LE} \leq 30$$

The second stage has optimization geometry selection. While being decoupled from each propulsion selection is not optimal, it makes the development of the interface easier. Owing to the small number of variables and ease of writing code, a particle swarm optimizer is used.

*3) Interface*

Python was chosen as the scripting language to carry out the optimization. The libraries/software used for the design all have a well exposed API which enables the use of already developed macros to carry out efficient numerical computation.

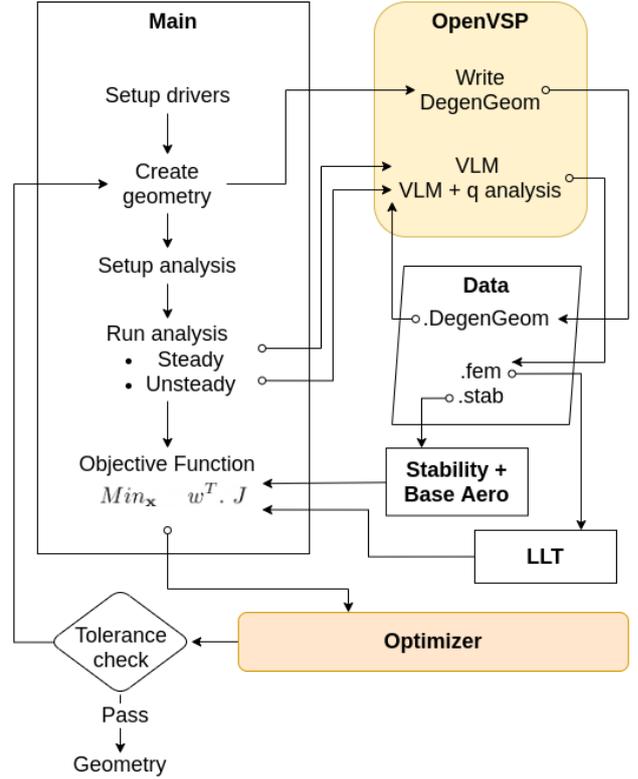

Fig 6. Preliminary Design Space

The geometry is arrived upon by interfacing a python script with OpenVSP and conducting a Vortex Lattice method analysis. The objective of the optimization accounts for three primary factors:

Drag: the analysis is done at constant lift coefficient which is equal to the lift at cruise condition.

Lift distribution; The lift distribution is a function of twist, taper ratio and sweep. A least squares cost function is used to match the target curve of the load distribution. A separate python script calculates this.

Stability: Since optimizing for stability often requires compromise on performance, the damping ratios of the short period, Dutch roll and phugoid mode are incorporated within the objective. For static stability however, a constraint for the design lift coefficient at zero moment is enough. The stability coefficients and BaseAero values are sought from the .stab file output by OpenVSP and converted to a data frame for further processing.

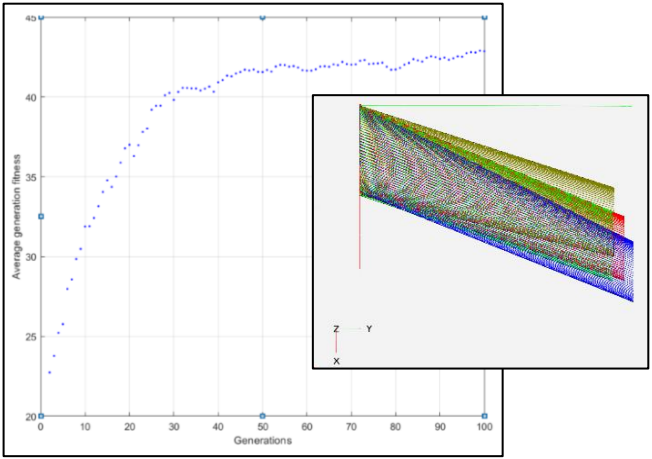

*Fig 7. Optimization tradeoffs(right), Avg. Fitness vs Generations(left)*

Some comments on each of design areas with respect to the tools are written to help the reader understand the level of sophistication available and possible caveats.

- *Vortex lattice method is not an accurate method for determining the overall drag and therefore the values are further verified by CFD analysis.*
- *The focus of the geometry optimization is primarily on the cruise phase.*
- *Since this is not a full-scale MDO solution, A human in the loop is necessary to guide and remove obviously suboptimal designs.*

Table 1.   AIRCRAFT PARAMETERS

| Parameter | Values |
|---|---|
| Wing loading | 164.88 N/m² |
| Power loading | 3.47 Kg/Kw |
| Wing Area | 0.418 m² |
| Aspect Ratio | 7.35 |
| Taper Ratio | 0.689 |
| Wing incidence | 5° |
| Relative Twist | -2 |
| Leading Edge sweep | 26.5° |

*C. Propulsion selection:*

To drive the aircraft with its airframe designed at its peak performance with reduced drag to achieve the most of endurance, a propulsion architecture was conceived.

*1) Forward Motor*

These motors were the most essential part in optimizing cruise segment of the flight. The right combination of propeller size, pitch and motor combination was combed through to achieve the peak optimized performance of the motors while cruising at 25 m/s. This was matched with the pitch speed of the motor operating on the thrust needed at cruise and hence maximizing endurance.

The initial design of our fixed wing yielded a cruise drag $C_{d\ cruise}$ whose drag would have to be overcome by over forward cruise motors.

Get KV

$$v_{pitch} = [x_{pitch} \times \omega_{cruise}]$$

$$KV_{cruise} = \left[\left(\frac{v_{pitch}}{x_{pitch}}\right) \times \frac{1}{V_p} \times 60 \times \emptyset_c\right]$$

$$\emptyset_c = \sqrt{\left(\frac{T_{max}}{D_{cruise}}\right)}$$

Subject to

$$v_{pitch} = 25 m/s$$

$$2.36 \leq \omega_{cruise} \leq 4.92$$

$$4" \leq x_{pitch} \leq 8"$$

$$0.5 \leq T_{max} \leq 0.8\ GTOW$$

$$D_{cruise} = \frac{1}{2}\rho v^2 S C_{d\ cruise}$$

$$V_p = 22.2V\ (6S\ Battery)$$

An RPM scaling factor $\emptyset_c$ is introduced keeping in mind the fact that the Required RPM increases quadratically with the required Thrust for a constant pitch propelled BLDC motor. This factor scales the motor RPM at cruise to what its RPM would likely be at Maximum of the basis of our requirements of $D_{cruise}$ and $T_{max}$.

BLDC motors operating at 22.2V (6S) battery packs were searched upon which could give the required thrust of 17N at 500m and a pitch speed of 25m/s. EMAX MT3515 650KV motor coupled with a 10"x6" propeller was finalized upon after iterations of various KV rating of motors.

The max load the motor could handle was 26N of thrust, comfortable for flying at max T/W of 0.74 and at *x + x + 2x* configuration of motor thrust distribution can easily hover with a thrust margin of 1.48:1 during hover segments.

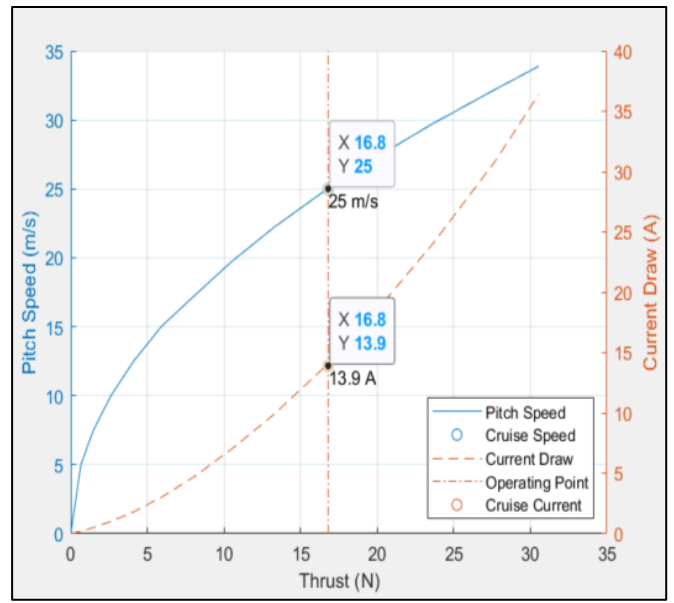

*Fig 8 Forward Motor Optimal Thrust*

## 2) Battery Selection

Observing modern trends in electric propulsion in aviation, Lithium Ion battery packs prove to be the best in terms of energy density, holding more amperage compared to their Lithium Polymer counterparts. Lithium Ion battery packs do have their drawbacks of a limited current discharge. Current was evaluated for all segments of flight and was found sufficient to use.

Table 2. BATTERY DATA

| Parameter | Values |
|---|---|
| Form factor | 18650 |
| Cell Voltage (nominal) | 22.2V |
| Cell Capacity | 3300mAh |
| Cell Discharge | 3C (9.9A) |
| Cell Discharge (Peak) | 4C (13.2A) |
| Weight (cell) | 48g |

$W_{battery}(\max) = 3 Kg$

$6S\ 10P = 2880g$

Hence for operating at a 6S 10P Configuration:

Flight range of more than 100Kms was subsequently verified by calculating the battery usage in cruised flight.

$$Flight\ Time = \frac{C_{cell} \times n_{Parallel\ cells}}{I_{cruise} \times n_{motors}} \times \eta_{total}$$

$Flight\ Time = 71\ mins$

$Range = v_{cruise} \times t_{cruise}$

$Range = 107\ Kms\ (Theoretical)$

## 3) Rear Motor

With the forward motors handling half of the thrust required during hover segments the remaining is handled by the rear motor. Its sole optimization goal is to draw the least amount of current so as to be as efficient as possible in doing its job. Motors with larger propeller size and lower RPM (KV rating) draw lesser current as the induced pitch speed is lesser than those with shorter propellers and higher RPM.

The rear motor was hence finalized by limiting the propeller diameter to the length of available clean area under the rear propeller wash disc keeping in mind the placement of the rear motor according to the CoG.

The collective propulsion results are validated on and electric propulsion calculator which is not free. The results are in close agreement with the developed formulae above.

Table 3. FINAL MOTORS SELECTION

| Forward Motor (×2) | Value |
|---|---|
| EMAX MT3515 | |
| Type | BLDC 22.2V |
| KV rated | 650 RPM/V |
| Maximum Power | 630 W |
| Propeller | 10" x 6" |
| Max Thrust | 2660g |
| Weight | 131g |
| **Rear Motor (×1)** | **Value** |
| T-MOTOR MN501-S | |
| Type | BLDC 22.2V |
| KV rated | 360 RPM/V |
| Maximum Power | 1000 W |
| Propeller | 20" x 6" |
| Max Thrust | 5372g |
| Weight | 171g |

## VI. DETAILED DESIGN

The preliminary design results in a low fidelity flyable aircraft capable of satisfying the theoretical performance objectives. More sophisticated software for Computational fluid dynamics and aircraft control simulation engines are used at this point to validate the design.

### A. CAD and preprocessing:

The solid model with all internal features is designed on FreeCAD. Extensive use of the part design and draft workbenches allows the creation of complex models and geometries. The commercial tool used for comparison is SolidWorks. As shown below, FreeCAD can capture most of the geometry but the tail is off design due to the lack of sophisticated lofting features with guide curves as compared to Solidworks.

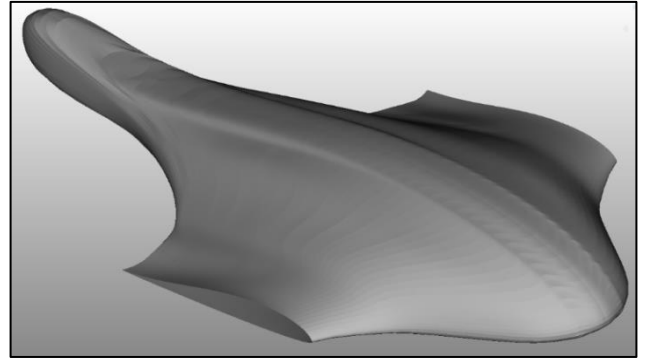

*Fig 9. FreeCAD Body*

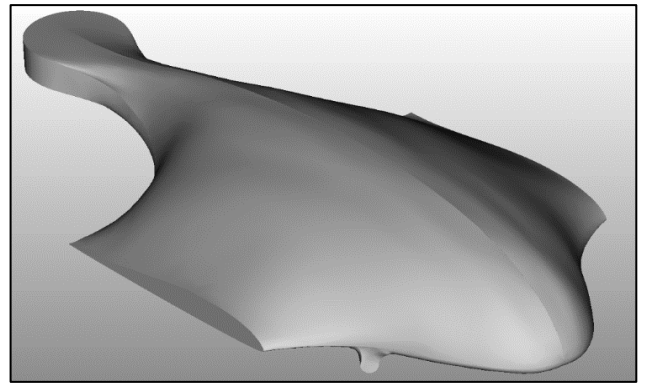

*Fig 10. SolidWorks Body*

Triangulation for the CFD model is done using the Draft, part and mesh design workbench. Each part and face are sequentially upgraded using the draft upgrade, fused using the boolean fuse option and finally the entire fusion is converted to a mesh. Owing to the tedious processing time of a Netgen mesh, the standard mesh was used and further mesh optimization and quality improvement was conducted in the CFD tool.

### B. CFD and postprocessing:

A computation fluid dynamics analysis was setup to calculate the lift and drag more accurately than the low fidelity analysis. OpenFOAM was used to do analysis.

For preprocessing, blockMesh and snappyHexMesh were used with the parameters given below. Some refinement regions were also added in particular areas like the leading edge and winglet-wing blend. The mesh quality dictionary was also edited to refine some areas. The following table shows only the edited values.

Table 4. MESH PARAMETERS

| Parameter | Value |
|---|---|
| addLayers | false |
| Leading edge refinement levels | (1E15 8) |
| Winglet refinement levels | (1E15 7) |
| Front motor refinement levels | (1E15 7) |
| Local refinement box levels | (1E15 5) |
| implicitFeatureSnap | false |
| mergeTolerance | 1e-6 |
| maxNonOrtho | 60 |
| minTwist | -1e+30 |
| minVol | -1e+30 |
| minDeterminant | -1 |
| minFaceWeight | -1 |
| minTriangleTwist | -1 |

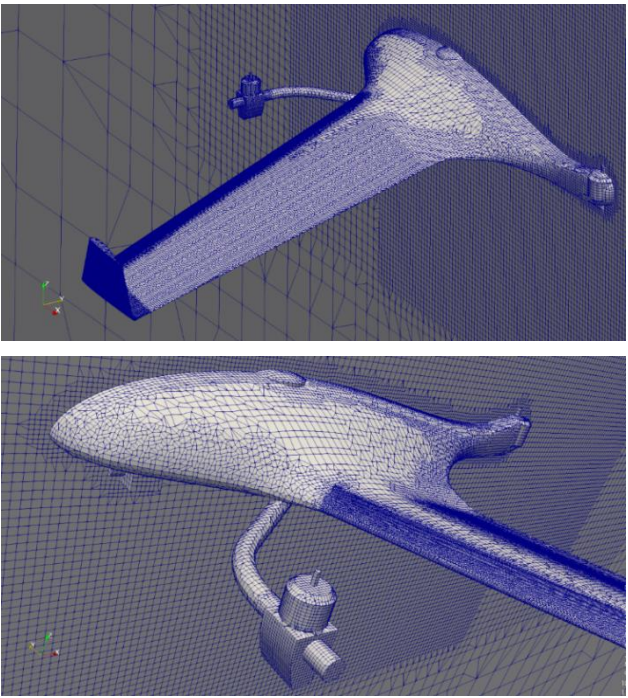

Fig 11. snappyHexMesh results

The basic steady state simpleFoam solver was used to obtain the coefficients at cruise state. The solver was run for 500 iterations but resulted in satisfactory residual convergence in 200 iterations (fig 13.)

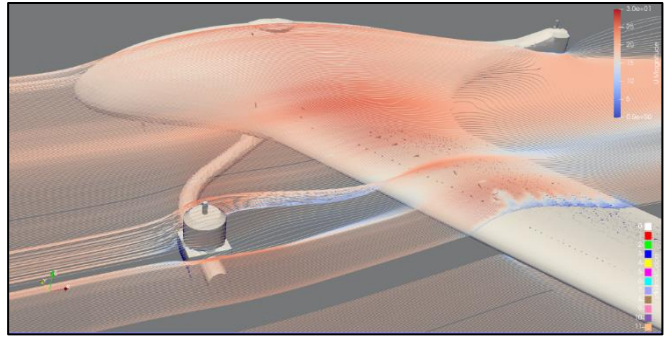

Fig 12. Streamlines coloured by velocity

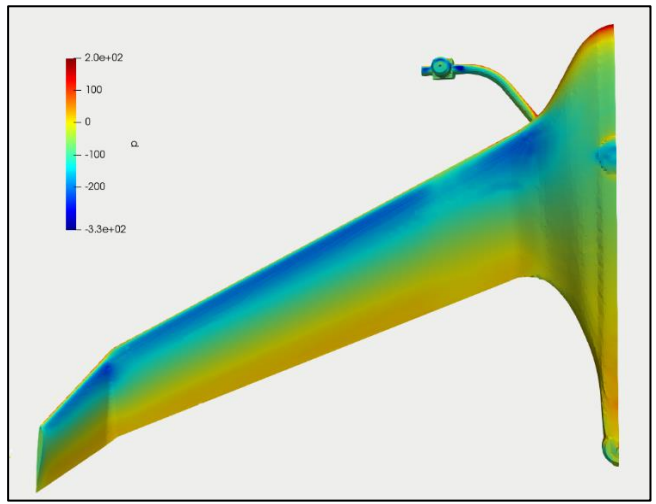

Fig 13. Pressure Contours

Table 5. OBTAINED COEFFICIENTS COMPARISON

| Coefficient | XFLR5 | OpenFOAM |
|---|---|---|
| $C_D$ | 0.018 | 0.038 |
| $C_L$ | 0.477 | 0.403 |

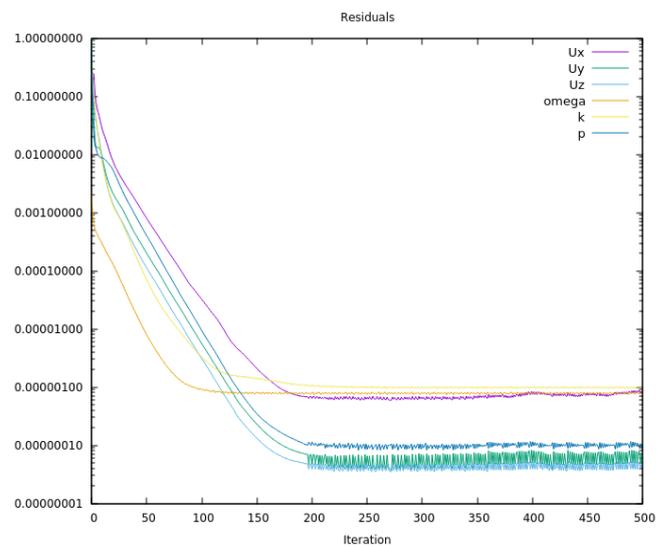

Fig 14. Residual convergence

## C. Render

While rendering is obviously not essential to the design process, it is required in some commercial sectors. Conventionally it is done using expensive tools and software. Contrarily, we have used LuxCoreRenderer as the opensource tool to do the same and included it in the design loop.

The raytracing module of FreeCAD is used to output a .lxs configuration file which is modified accordingly. Note that the default .lxs file template has preset sizes and transformations of area light sources and one can either change the config file or the CAD itself. Our option was to choose the former which again highlights the freedom offered by opensource software. The changes to the default configuration are given. The entire render with a final noise pass error of about 0.001 took about 15 minutes to complete.

```
1. Camera
2. LookAt
       42.9652           27.7823           39.252
       -7.02438          -22.438           -0.599087
       -0.336994         -0.35615          0.871547
3. Flex image
4. "integer xresolution" [1080]
5. "integer yresolution" [1360]
6. Groundplane
7. Transform
       [4.000000000000000     0.000000000000000
       0.000000000000000      0.000000000000000
       0.000000000000000      4.000000000000000
       0.000000000000000      0.000000000000000
       0.000000000000000      0.000000000000000
       4.000000000000000      0.000000000000000
       0.000000000000000      0.000000000000000
       0.000000000000000      4.000000000000000]
8. Area Lightsource : sun
9. Transform
       [15.000000000000000    0.000000000000000
       0.000000000000000      0.000000000000000
       0.000000000000000      15.000000000000000
       0.000000000000000      0.000000000000000
       0.000000000000000      0.000000000000000
       15.000000000000000     0.000000000000000
       0.000000000000000      0.000000000000000
       40.000000000000000     1.000000000000000]
10."float power" [200.000000000000000]
11."color L" [0.50000000 0.95364237 0.70636380]
```

A comparison between Solidworks PhotoView and LuxCore is also done:

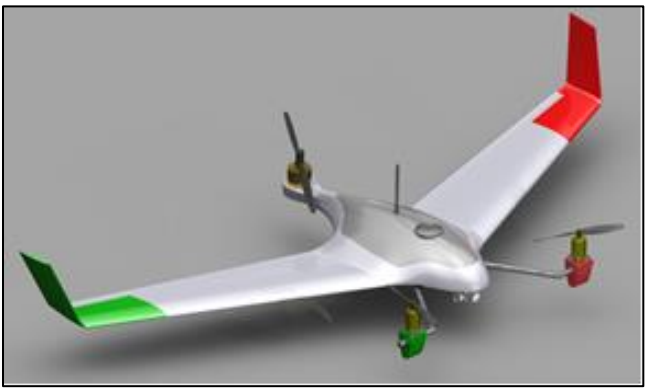

Fig 15. Solidworks+PhotoView Render (top)

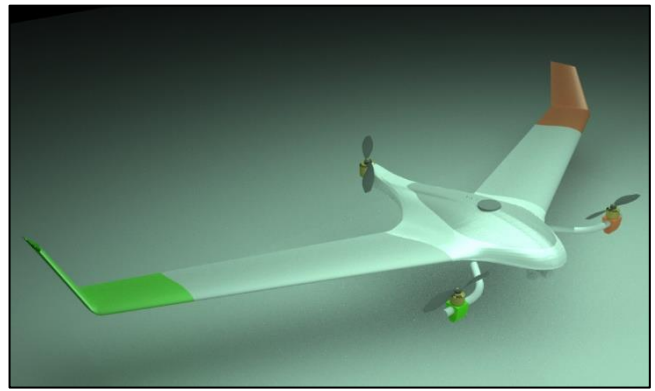

Fig 16. FreeCAD + LuxCore Render (bottom)

VII. DEVELOPMENT AND INTEGRATION

## A. Fabrication:

A prototype body was fabricated to validate some performance and stability characteristics of the aircraft. The body is made of Extruded polystyrene foam (XPS) and glass fiber fabric through a vacuum assisted wet layup method. Some aspects of the configuration were altered depending on the resources available.

The fuselage was made by hotwiring along each orthographic projection plane. This method was highly efficient in terms of time and ensured minimum sanding of the foam to get the final shape. Aluminum pipes were used to define the structure of the tricopter assembly and the foam blended wing body was integrated along.

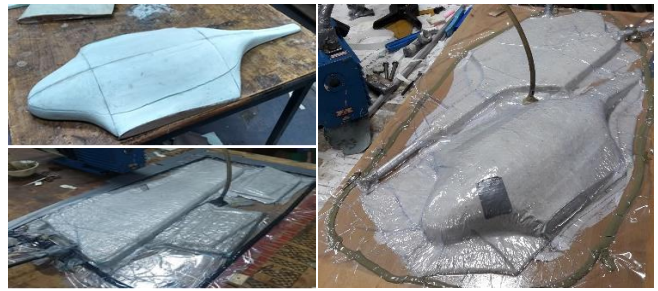

Fig 17. Fabrication and Vaccum bagging

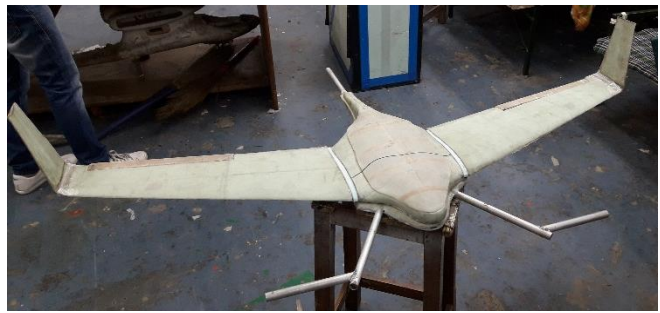

Fig 18. Assembled body (1st iteration)

## B. Integration

A well-defined architecture is shown in the fig below highlighting data flow between the subsystems.

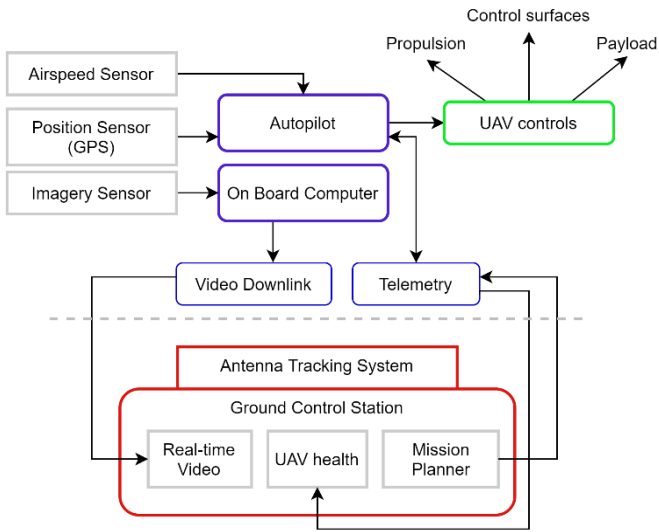

*Fig 19. System Architecture*

A brief description of some of the components is given:

*1) Pixhawk 2 Cube:* An autopilot is necessary to carry out custom missions and provide additional stability during flight. It also allows the UAV to maintain a level altitude while transitioning into cruise state.

*2) GPS:* A Here GPS is used to localize the UAV in its environment.

*3) Telemetry:* An RFD 900x telemetry is used to create a link between the ground systems and send appropriate messages in between the flight.

*4) Sensors:* These include everything ranging from imaging systems, airspeed sensors and

*5) Raspberry Pi 3:* The onboard computer's primary purpose is, to carry out the image recognition task among other necessary mission scripts.

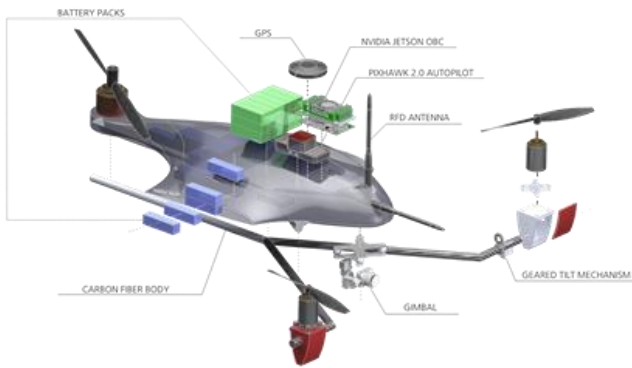

*Fig 20. Exploded View of subsystems*

A key feature is the form of the modular propulsion batteries. Multiple Lithium Ion cells can be used to form arrangements so as to fit into the body. The blended design offers a lot of payload volume which is exploited by stacking 9 parallel sets of 6 cells each connected in series. Some of the battery packs are strategically placed very close to the CoG which allows the user to swap the pack for additional payload or sensors.

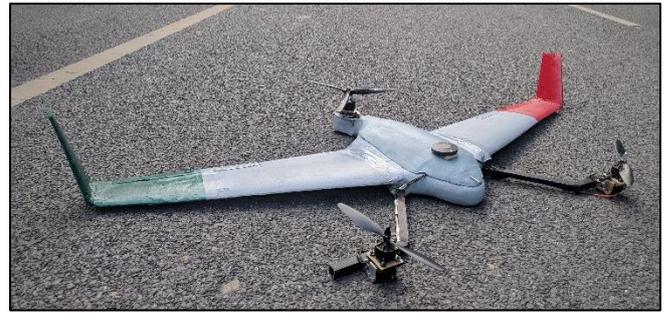

*Fig 21. Final RTF prototype*

### C. Flight Testing:

Table 6. TUNED FLIGHT PARAMETERS

| Parameter | Original | Changed |
|---|---|---|
| ARSPD_FBW_MAX | 30 | 28 |
| ARSPD_FBW_MIN | 10 | 16 |
| PTCH2SRV_D | 0.200000003 | 0.04 |
| PTCH2SRV_P | 2.5 | 1 |
| Q_FRAME_CLASS | 1 | 7 |
| Q_TILT_YAW_ANGLE | 0 | 30 |
| Q_A_RAT_PIT_P | 0.25 | 0.2 |
| Q_A_RAT_YAW_P | 0.1800000072 | 0.25 |
| Q_A_THR_MIX_MAN | 0.1000000015 | 0.5 |

The prototype was tested physically over a few flight tests to evaluate flight dynamics and tune the autopilot. Mission Planner was used for tuning along with ArduPlane software. The testing resulted in many unforeseen issues which were resolved after each iteration. Some key points of discussion are noted:

*1)* The low cost and quality of materials due to the tight budget made it very difficult to debug what the problem was and resulted in increased number of flight times.

*2)* Lack of prior experience and exposure in handling or operating a hybrid craft made every challenge as a new problem to be tackled.

*3)* Even post debugging the problem, finding the right solution was a tedious task due to the vast number of PID parameters to tune. A great help at this time was the shared experience of the community at the arduplane forums.

## VIII. CONCLUSION

A VTOL unmanned aerial system capable of flying 100km and performing a relief mission is developed. Particular care was taken that no commercial tools or codes are used. Using completely opensource tools has its own sets of challenges and advantages. Some comparisons and discussion are therefore required to close the loop.

### A. Discussion

Aircraft design is and always will be a multi-processing task where no linear approach exists and no defined starting point awaits. Even a piece of cardboard is airworthy enough to fly, but for it to take shape into an optimized craft with required capabilities, various parameters have to be tweaked simultaneously with one affecting the other at every step in the design process. To overcome such complexities, communities exist online as well in laboratories, aiming to pool in the unquantifiable amount of knowledge of Aero Design into open source software and libraries:

*1) Opensource tools are not without their limitations. A significant effort was required to get acquainted with all tools at every stage and produce meaningful results.*

*2) A key difference which tends to take up significant time is the lack of standardised coordinate systems, conventions of unit systems, and general understanding of some concepts between each softwrae. This is because opensource softwares, while guided by some core teams are the heart of it-collaborative. And this results in different conventions on each platform. Effort from the user's side is required to understand this point and change models/configurations accordingly*

*3) The above point becomes more important when there is significant lack of documentation, discussion forums and in general people using it. To overcome this , the clarity of fundamentals of types software architectures and the core field are important from the user's side.*

*4) The issue of 'accuracy' of results was verified collectively by reading up on literature and conducting some of the analyses on commercial software and calculating errors.*

ACKNOWLEDGMENT

The UAV was fabricated at Team UAS-DTU laboratory, Delhi Technological University and the authors duly acknowledge the help of the team members for the same.